# Patent Citation Spectroscopy (PCS):

## Algorithmic retrieval of landmark patents

Jordan A. Comins[1,*,‡], Stephanie A. Carmack[2], and Loet Leydesdorff[3]

**Abstract**

One essential component in the construction of patent landscapes in biomedical research and development (R&D) is identifying the most seminal patents. Hitherto, the identification of seminal patents required subject matter experts within biomedical areas. In this brief communication, we report an analytical method and tool, Patent Citation Spectroscopy (PCS), for rapidly identifying landmark patents in user-specified areas of biomedical innovation. PCS mines the cited references within large sets of patents and provides an estimate of the most historically impactful prior work. The efficacy of PCS is shown in two case studies of biomedical innovation with clinical relevance: (1) RNA interference and (2) cholesterol. PCS mined and analyzed 4,065 cited references related to patents on RNA interference and correctly identified the foundational patent of this technology, as independently reported by subject matter experts on RNAi intellectual property. Secondly, PCS was applied to a broad set of patents dealing with cholesterol – a case study chosen to reflect a more general, as opposed to expert, patent search query. PCS mined through 11,326 cited references and identified the seminal patent as that for Lipitor, the groundbreaking medication for treating high cholesterol as well as the pair of patents underlying Repatha. These cases suggest that PCS provides a useful method for identifying seminal patents in areas of biomedical innovation and therapeutics. The interactive tool is free-to-use at: www.leydesdorff.net/pcs/.

---

[1] *corresponding author*; Social and Behavioral Sciences Department, The MITRE Corporation, McLean, VA, United States; jcomins@gmail.com

[‡] The author's affiliation with The MITRE Corporation is provided for identification purposes only, and is not intended to convey or imply MITRE's concurrence with, or support for, the positions, opinions or viewpoints expressed by the author. Approved for Public Release; Distribution Unlimited Case #17-0951.

[2] National Institute on Drug Abuse, National Institutes of Health, Baltimore, MD, 21224

[3] Amsterdam School of Communication Research (ASCoR), University of Amsterdam, PO Box 15793, 1001 NG Amsterdam, The Netherlands

**Introduction**

Amongst the various components of a patent landscape, identifying seminal patents in an innovation area requires substantial investment from specialists (e.g., Schmidt, 2007). Hitherto, , subject matter experts review a large corpus of patents and patent applications within their historical context to render a judgment of the most technologically important patents. This method is time-consuming, difficult to replicate, and predicated on the availability of subject matter experts (Cockburn et al., 2002) – and yet, there is a requirement for patent examiners and historians of science. Thus, there is a need for automated methods for uncovering insights about landmark patents in technology areas (Jensen and Murray, 2005; Konski and Spielthenner, 2009).

Clinical advances depend upon a sound understanding of biomedical research and development. The importance of maintaining situational awareness of biomedical R&D activities for businesses and policy-makers is best exemplified by the proliferation of patent landscapes produced by subject matter experts covering a wide range of topics (e.g., CRISPR: Egelie et al, 2016; Induced Pluripotent Stem Cells: Roberts et al, 2014, Bergman & Graff, 2007; Prenatal Testing: Agarwal et al., 2013; Carbon Nanotubes: Harris and Bawa, 2007; Nanomedicine: Wagner et al, 2006; Gene Sequences: Jensen & Murray, 2005). Given the enormous growth in the number of annual patent applications filed (USPTO, 2016), particularly in the life and biomedical sciences (Moses et al., 2015), there is increasing demand for patent landscapes across a panoply of technologies. In light of this growing need, we introduce an algorithmic approach and web-application for identifying landmark patents, a key component of patent landscapes, across user-specific biomedical areas.

Patent Citation Spectroscopy (PCS) operates over the cited references of large sets of patents to determine the seminal prior works within a given field, as well as an openly available web-application for performing PCS. In this brief communication, we apply PCS to two areas of biomedical innovation: (1) RNA interference and (2) cholesterol. RNAi was selected to examine the efficacy of PCS for understanding the origins of an emerging technology with well-documented expert reviews to ground our findings (Schmidt, 2007). Cholesterol was selected to consider how PCS performs for searches on broad areas of biomedical innovation and clinical relevance that reflect the less sophisticated kind of searches conducted by users that are not library scientists or patent experts.

**Methods**

**PCS Algorithm**

PCS can be performed over a set of US patent data that includes a list of referenced patents, or backward citations. Our study utilizes data from PatentsView, a data platform sponsored by the USPTO. PatentsView provides backward citation information for all US patents from 1976 through July 2016 via an Application Program Interface (API). We leverage this API both in demonstrating the utility of PCS to identify seminal patents and in creating a tool that makes PCS widely available for further use.

The core set of computations performed by PCS are largely consistent with that of the Reference Publication Year Spectroscopy (RPYS) technique (Marx et al., 2014), with one notable exception: the normalization is different. Briefly, RPYS aggregates the references within documents and organizes those references by their publication date. Once organized, these values are summed so that each reference publication year represents the cumulative number of references it received from the citing set of documents. Then, for each reference publication year, the absolute deviation between the focal year and the median number of references for the five-year period that includes the focal year is calculated. The resulting vector is traditionally graphed as a line. Viewing the line, analysts then search for maxima, or peaks, as indicators of seminal works being published in a given year. Thus far, RPYS has successfully been used to identify seminal research publications across a multitude of scientific domains (Comins and Hussey 2015a; Elango *et al.* 2016; Leydesdorff *et al.* 2016; Leydesdorff *et al.* 2014; Marx and Bornmann 2013; Marx *et al.* 2014; Thor *et al.* 2016; Wray and Bornmann 2014; Comins and Hussey 2015b; Comins and Leydesdorff 2016). In addition, recent work shows the convergence between RPYS and subject matter expert's identification of seminal scientific papers within given areas of basic biomedical research (Comins & Leydesdorff, 2017).

Returning to the case of PCS, the relevant patents are first retrieved from the PatentsView API. Next, the cited patent references within each patent of this relevant result set are extracted and organized by the year those references were granted. We then sum the number of references attributed to each year. These data are de-trended by taking the absolute deviation of the number of cited references for a given reference year from the 5-year median of patent references. This is represented by the equation:

$$f(t) = C_t - Md(C_{t-2}, C_{t-1}, C_t, C_{t+1}, C_{t+2}) \qquad (1)$$

where *C* represents the total sum of citations to patents granted in year *t* and *Md* represents the median. These first steps mirror those found in RPYS.

In contrast to RPYS, PCS contains an additional normalization step. Specifically, the de-trending function discussed thus far only considers the aggregated cited reference activity over time. This can be problematic for finding seminal works as a very high "peak" of an RPYS de-trending function could be due to either a large surge in the influence of a single document (i.e., a seminal work) or due to the influence of multiple documents in a given year. To help us isolate the former cases, we extend the de-trending procedure as follows:

$$PCS(t) = f(t) \cdot \frac{\text{Count of References to Most Referenced Patent in Year } t}{C_t} \qquad (2)$$

Thus, by Equation 1 years with abnormally high numbers of references are identified, and by Equation 2 the values from Equation 1 are reduced based on the percentage of all references from that year coming from the most referenced patent.

**PCS Application**

Figure 1 shows the web-application implementing the PCS analysis for open use (hosted at www.leydesdorff.net/pcs). The application takes from the user a keyword (e.g., RNAi) or phrase query (e.g., "interference RNA"). Searches can be combined by the OR operator through the inclusion of commas (e.g., *RNAi,*"*interference RNA*", *siRNA*, *"RNA interference"*). Once a user clicks the *Search* button (Fig. 1, *upper left*), a GET request is sent to the PatentsView API (see Methods for details) to return metadata on any granted US patents containing those keywords or phrases in the title or abstract. The returned metadata arrives in the form Javascript Object Notation (json) with the relevant fields for conducting PCS (e.g., cited patents, cited patents' granted dates). Data is visualized using the HighCharts JavaScript library, which is free to use for non-commercial purposes.

The application provides two pieces of summary data (Fig. 1, *upper right*). The first is the number of granted US patents in the PatentsView database that were returned from the user's search. The second is the number of unique patents referenced by the cohort of patents from the search results.

The main output of the PCS tool is the reference spectrum of these returned patents revealing the backward citation distribution. Within the plot, two visualizations are rendered. The first is the raw count of backward citations arranged by the granted year of the references (Fig. 1, *left*). This is represented via a bar graph corresponding to the left y-axis. The second is a smoothed lline of the normalized score of backward citations arranged by the granted year of the references. This graph aligns with the right y-axis (Fig. 1, *right*).

The visualization is interactive, allowing users to explore results. Specifically, when the user's mouse hovers over data corresponding to a given reference granted year, a tooltip provides the value of both the left and right y-axis. In addition, clicking on any reference year will redirect the user to the full patent document corresponding to the most referenced patent granted that year. Finally, the system determines which referenced patent has the highest positive value in its normalized citation score and tells the user in plain text that this is the tool's estimation of the most impactful historical patent for the corresponding search (Fig. 1, *bottom*).

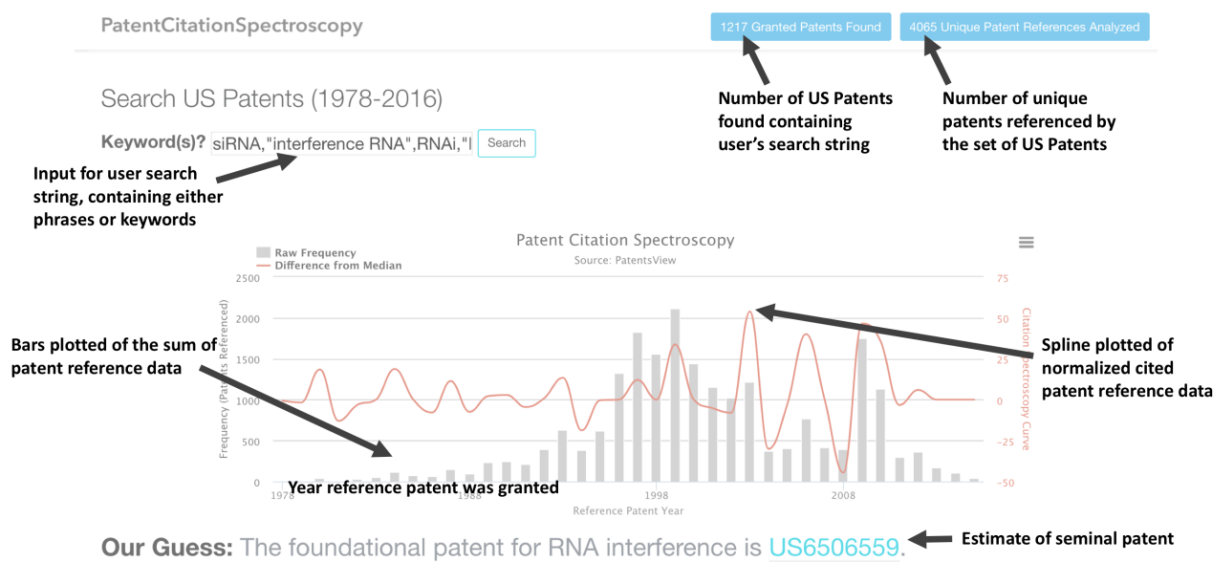

**Figure 1**. Image of the PCS web-application. In this demo, the user queried patents containing either the key terms "RNAi" or "siRNA" or the phrases "interference RNA" or "RNA interference." The system then searched the titles and abstracts of US patents within the PatentsView database for these search terms. The result was 1,217 granted US patents containing 4,065 unique patent references. The patent references were analyzed via PCS to produce a visualization of the spectrum of impactful historical patent references. PCS identified the most important historical patent for this field: US6506559 – Genetic inhibition by double-stranded RNA by Fire et al. (2003), a finding that converges with independent reports from subject matter experts (Schmidt et al., 2007).

**Results**

**Case Study 1: RNAi, an emerging technology area**

To demonstrate the utility of this tool, we applied PCS to an area of biomedical innovation: RNA interference. We selected RNA interference (hereto RNAi) as a use case for two reasons: (1)

RNAi represents a burgeoning domain of biomedical innovation with potential therapeutic applications for the treatment of viral infections and cancer; and (2) the patent landscape of RNAi has been studied by subject matter experts, which allows us to compare the results of PCS with their conclusions (Schmidt, 2007).

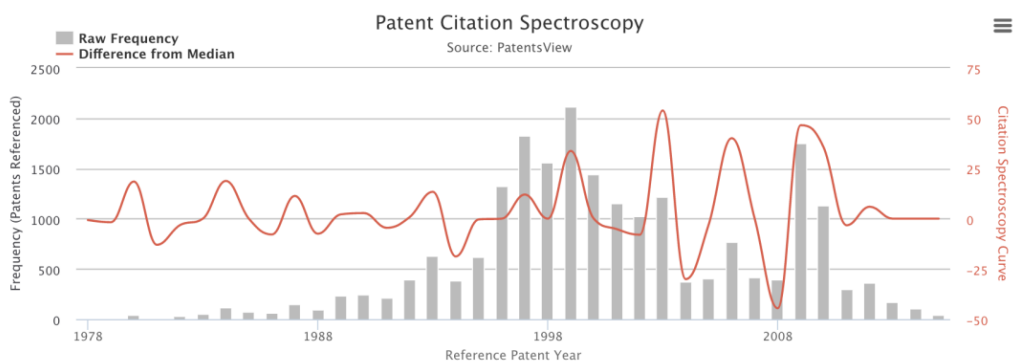

**Figure 2**. PCS plot for RNAi. The analysis identified the single most impactful patent in this area of innovation to be US6506559 by Fire et al granted in 2003, converging with the opinion of a subject matter expert. The odds of correctly identifying this patent by chance were 0.00025.

In conducting our analysis on RNAi, we chose as our search query terms that were previously used to explore RNAi from a scientometric perspective (Leydesdorff & Rafols, 2011; Rotolo et al., 2017). Specifically, we input either the strings *siRNA* or *RNAi* or *interference RNA* or *RNA interference* and the tool searched for US patents containing these terms in the title or abstract. As of FEB2017, this search yields 1,218 granted US patents possessing 4,065 unique patent references. Given the number of references analyzed, the odds of correctly guessing the key seminal patent by chance are 1 out of 4,065, or 0.00025. Conducting PCS on this search string, the tool produced US6506559 – *Genetic inhibition by double-stranded DNA* by Andrew Fire and colleagues, which was granted in 2003, as its estimate of the most seminal patent. The patent, commonly referred to as the Carnegie patent for the assignee organization, is clearly described as the foundational patent for RNAi (Schmidt, 2007). The binomial test of selecting the foundational patent reference correctly on the first attempt is significant ($P = 0.0005$).

We note that if running this analysis without the addition of our new normalization procedure, the highest peak would be attributed to 2009 (based on the traditional RPYS process). To users, this would suggest the most significant historical patent for RNAi occurred in 2009 and likely attributable to that year's most cited work, a patent by Leake et al (US7595387). However, this result would be inconsistent with the foundational patent identified by subject matter experts and shows that the absence of this new normalization procedure makes it more difficult for the user to veridically ascertain the most seminal patent in a field.

Without the new normalization procedure, peak height in the resulting graph is not influenced by the impact of a given year's most referenced work. Our results, on the other hand, highlight the value of the new normalization procedure of PCS, which appropriately shifts the highest peak from 2009 to 2003 due to the large share of cited references in 2003 pointing to the patent by Fire et al. The new normalization procedure thereby simplifies the task of identifying the most seminal patent in the field of RNAi for the user.

In addition to correctly specifying the foundational patent for RNAi, the PCS application identified several other patent references of interest (as shown by peaks in the graph). Of particular note, the peak observed in 2006 corresponds to US7056704 – *RNA interference mediating small RNA molecules* by Thomas Tuschl and colleagues. Schmidt (2007) includes this patent amongst the most seminal in the RNAi field as it details siRNA architectures thought to be critical drug efficacy.

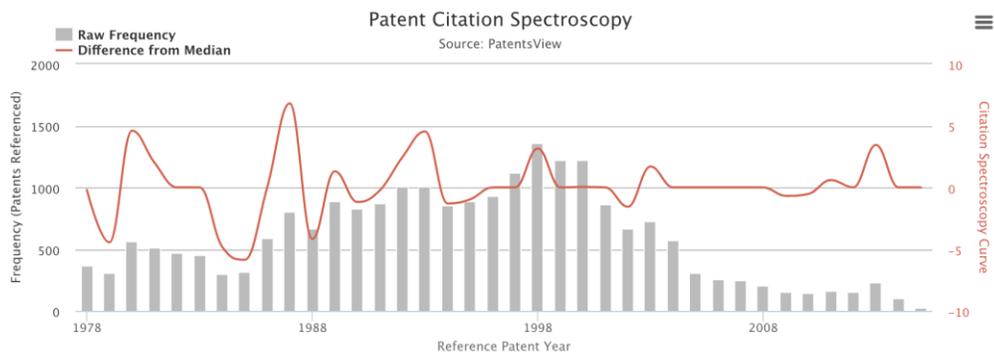

**Figure 3**. PCS plot for cholesterol. The patent identified as most seminal in this area was US4681893 by Bruce Roth granted in 1987. This represents the fundamental patent underlying the medication Atorvastatin (ie., Lipitor).

**Casse Study 2: Cholesterol, a broad area of biomedical importance**

In our second case study, we opted to conduct PCS on a broad area of biomedical relevance. Thus, for this case, we searched for the term *cholesterol* with the aim of uncovering the seminal patent pertaining to cholesterol therapies or measurement. The PCS method identified US4681893, granted in 1987, as the most seminal patent. This patent, produced by Bruce Roth at the Warner-Lambert Company, is the foundational intellectual property underlying Atorvastatin (Hutchins, 2003). Otherwise known as Lipitor, this technology has played a tremendous role in preventing cardiovascular disease. Though it expired in 2009, this patent played a transformative role for biomedical innovations dealing with cholesterol. Also of note in the PCS plot are a pair of peaks corresponding to patents US8030457 granted in 2011 and US8563698 granted in 2013, both by Simon Jackson and colleagues. These patents are central to the production of evolocumab. More commonly known by the brand name Repatha, evolocumab is an antibody that is used to treat hypercholesterolemia (Markham, 2015).

**Discussion**

Identifying intellectual pathways within biomedical science and technology is an important component of patent landscapes required by businesses and policy-makers. The rapid identification of landmark patents by PCS will support the generation of data-driven patent landscapes. Using the PCS methodology and application described here, it will be easier for decision-makers to understand the fundamental patents of myriad biotechnologies as well as the companies (assignees) and people (inventors) responsible for them.

To demonstrate the effectiveness of PCS, we applied the application to two distinct use-cases. The first is for RNA interference, an emerging technology area with the potential for new therapeutic interventions. Compared with prior subject matter expert reports on RNAi patents (Schmidt, 2007), PCS correctly identified the foundational patent by Andrew Fire et al (2003). Notably, lead authors Andrew Fire and Craig Mello on this seminal patent were both awarded the Nobel Prize for their research on RNAi. Our second case study focused on cholesterol, a broader topic area of biomedical relevance. For this area, PCS identified the patents underlying Lipitor and Repatha, two critical biomedical therapies for maintaining healthy levels of cholesterol. In short, PCS provides research policy makers, life scientists and scientometricians broader access to a tool capable of identifying the important historical influences in areas of innovation.